\documentclass[12pt,thmsa]{article}%
\usepackage{amsmath}
\usepackage{amssymb}
\usepackage{amsfonts}
\usepackage{graphicx}%
\begin{document}

\title{Adiabatic passage in a three-state system with non-Markovian relaxation: The
role of excited-state absorption and two-exciton processes.}
\author{B. D. Fainberg$^{1,2}$\thanks{Corresponding author. E-mail:
fainberg@hit.ac.il} and V. A.Gorbunov$^{1}$\\$^{1}$Faculty of Sciences, Physics Department, Holon Institute of Technology, \\52 Golomb St., Holon 58102, Israel\\$^{2}$Raymond and Beverly Sackler Faculty of Exact Sciences, \\School of Chemistry, Tel-Aviv University, Tel-Aviv 69978, Israel}
\date{\today }
\maketitle

\begin{abstract}
The influence of excited-state absorption (ESA) and two-exciton processes on a
coherent population transfer with intense ultrashort chirped pulses in
molecular systems in solution has been studied. An unified treatment of
adiabatic rapid passage (ARP) in such systems has been developed using{ a
three-state electronic system with relaxation treated as a diffusion on
electronic potential energy surfaces. }We have shown that ESA has a profound
effect on coherent population transfer in large molecules that necessitates a
more accurate interpretation of experimental data. A simple and physically
clear model for ARP in molecules with three electronic states in solution has
been developed by extending the Landau-Zener calculations putting in a third
level to random crossing of levels. A method for quantum control of
two-exciton states in molecular complexes has been proposed.

\end{abstract}

\section{Introduction.}

The possibility of the optical control of molecular dynamics using properly
tailored pulses has been the subject of intensive studies in the last few
years
\cite{Ruh90,Kra93,Koh95,Mel91,Bar95,Gar95,Nib92PRL,Dup93,Ste96,Hil96,Cer96,Fai98,Mis98,Bar99,Fai00JCP,Mis99,Lin99,Mis00,Kal00,Manz00,Mal01,Gelman_Kosloff05,Sension06JPC}%
. Chirped pulses can selectively excite coherent wave packet motion either on
the ground electronic potential energy surface of a molecule or on the excited
electronic potential energy surface due to the intrapulse pump-dump process
\cite{Ruh90,Bar95,Cer96,Fai98}. In addition, they are very efficient for
achieving optical population transfer between molecular electronic states.
Total electronic population inversion can be achieved using coherent
light-matter interactions like adiabatic rapid passage (ARP) in a two- or
three-state system \cite{Ber98,Vit01}, which is based on sweeping the pulse
frequency through a resonance.

Since the overwhelming majority of chemical reactions are carried out in
liquid solution, adiabatic passage in molecules in solution was studied for
two-state electronic system (ARP) in Refs.\cite{Fai02JCP_2,Fai04JCP,Nag02},
and for stimulated Raman adiabatic passage (STIRAP)\ configuration in
Refs.\cite{Rice02,Geva03}. It has been shown in Ref.\cite{Fai02JCP_2} that
relaxation does not hinder a coherent population transfer for positive chirped
pulses and moderate detuning of the central pulse frequency with respect to
the frequency of Franck-Condon transition.

However, a two electronic state model for molecular systems is of limited
utility. Indeed, excited-state absorption (ESA) occurs for majority of complex
organic molecules \cite{Kov97,Sension06JPC}. Even a molecular dimer consisting
from two-level chromophores has an additional excited state corresponding to
two-exciton excitation. An unified treatment of ARP in such systems can be
developed using three-state electronic system interacting with reservoir (the
vibrational subsystems of a molecule (chromophores) and a solvent).

More often than not ESA in complex organic molecules corresponds to a
transition from the first excited singlet state $S_{1}$ to a higher singlet
state $S_{n}$ ($n>1$), which relaxes back to $S_{1}$ very \ fast
\cite{Bog78,Bog78-2,Bog82,Kov97}. Therefore, it would look as if ESA does not
influence on population transfer $S_{0}$ $\rightarrow$ $S_{1}$ from the ground
state $S_{0}$. However, in the presence of ESA an exciting pulse interacts
with both $S_{0}$ $\rightarrow$ $S_{1}$ and $S_{1}$ $\rightarrow$ $S_{n}$
transitions. It is well known that coherent optical interactions occurring in
adjacent optical transitions in a three-state system markedly affect each
other. The examples are STIRAP, lasing without inversion, coherent trapping,
electromagnetically induced transparency and others. (For textbook treatments
of these effects see, for example, \cite{scully-zubairy.97}). Therefore, one
would expect an appreciable change of a population transfer $S_{0}$
$\rightarrow$ $S_{1}$ with chirped pulses in the presence of excited state
absorption in the coherent regime when the chirp rate in the frequency domain
is not large and, consequently, the pulse is rather short.

Our objective is to answer the following questions: \textquotedblleft How do
ESA and two-exciton processes influence on a coherent population transfer in
molecular systems in solution? What is the potential of chirped pulses for
selective excitation of the single and two-exciton states and their selective
spectroscopy?\textquotedblright

In addition, the three-state system under discussion enables us to consider
STIRAP as well. Therefore, we shall also briefly concern slowing down the pure
dephasing on STIRAP in intense fields when relaxation is non-Markovian.

The outline of the paper is as follows. In Sec.\ref{sec:equations} we present
equations for the density matrix of a three-state molecular system under the
action of shaped pulses when the interaction with a dissipative environment
can be described as the Gaussian-Markovian modulation (so called the total
model). In Sec.\ref{sec:approx_models} we formulate a number of approaches to
this model that enables us, first, to clarify the underlying physics and,
second, to understand the validity of the results obtained by the total model.
The ESA effects on ARP in complex molecules are considered in
Sec.\ref{sec:adiabatic_pop_transfer}. In Sec.\ref{sec:excitons} we study
population transfer in molecular dimers with taking into account two-exciton
processes. In Sec.\ref{sec:STIRAP} we consider slowing down the pure dephasing
on STIRAP in strong fields when the system-bath interaction is not weak
(non-Markovian relaxation). We summarize our results in
Sec.\ref{sec:conclusion}. In the Appendix we extend calculations of two-photon
excitation of a quantum ladder system by a chirped pulse \cite{Gir03} to
non-zero two-photon detuning.

\section{Basic equations}

\label{sec:equations}

Let us consider a molecular system with three electronic states $n=1,2$ and
$3$ in a solvent described by the Hamiltonian
\begin{equation}
H_{0}=\sum_{n=1}^{3}|n\rangle\left[  E_{n}+W_{n}(\mathbf{Q})\right]  \langle
n| \label{eq:hamilt}%
\end{equation}
where $E_{3}>E_{2}>E_{1},$ $E_{n}$ is the energy of state $n,W_{n}%
(\mathbf{Q})$ is the adiabatic Hamiltonian of reservoir $R$ (the vibrational
subsystems of a molecular system and a solvent interacting with the
three-level electron system under consideration in state $n$).

The molecular system is affected by two shaped pulses of carrier frequencies
$\omega_{1}$ and $\omega_{2}$
\begin{equation}
\mathbf{E}(t)=\frac{1}{2}\sum_{i=1,2}\mathbf{E}_{i}(t)+c.c.=\frac{1}{2}%
\sum_{i=1,2}\mathcal{\vec{E}}_{i}\left(  t\right)  \exp[-i\omega_{i}%
t+i\varphi_{i}\left(  t\right)  ]+c.c. \label{eq:field}%
\end{equation}
which are resonant to optical transitions $1\rightarrow2$ and $2\rightarrow3$,
respectively (ladder configuration). Here $\mathcal{E}_{i}\left(  t\right)  $
and $\varphi_{i}\left(  t\right)  $ describe the change of the pulse amplitude
and phase, respectively, in a time $t$. The instantaneous pulse frequencies
are $\omega_{i}\left(  t\right)  =\omega_{i}-\frac{d\varphi_{i}}{dt}$ .

The influence of the vibrational subsystems of a solute and a solvent on the
electronic transition can be described as a modulation of this transition by
low frequency (LF) vibrations $\{\omega_{s}\}$ \cite{Fai80,Fai87}. In
accordance with the Franck-Condon principle, an electronic transition takes
place at a fixed nuclear configuration. Therefore, for example, the quantity
$u(\mathbf{Q})=W_{2}(\mathbf{Q})-W_{1}(\mathbf{Q})-\langle W_{2}%
(\mathbf{Q})-W_{1}(\mathbf{Q})\rangle_{1}$ is the disturbance of nuclear
motion under electronic transition $1\rightarrow2$. Here $\langle\rangle
_{n}\equiv Tr_{R}\left(  ...\rho_{R_{n}}\right)  $ denotes the trace operation
over the reservoir variables in the electronic state $n$,

$\rho_{R_{n}}=\exp\left(  -\beta W_{n}\right)  /Tr_{R}\exp\left(  -\beta
W_{n}\right)  ,$ $\beta=1/k_{B}T$.

The relaxation of electronic transition $1\rightarrow2$ stimulated by LF
vibrations is described by the correlation function $K(t)=\langle
u(0)u(t)\rangle$ of the corresponding vibrational disturbance with
characteristic attenuation time $\tau_{s}$ \cite{Fai98,Fai87}. We suppose that
$\hbar\omega_{s}\ll k_{B}T$. Thus $\{\omega_{s}\}$ is an almost classical
system and operators $W_{n}$ are assumed to be stochastic functions of time in
the Heisenberg representation. The quantity $u$ can be considered as a
stochastic Gaussian variable. We consider the Gaussian-Markovian process when
$K(t)/K(0)\equiv S(t)=\exp(-|t|/\tau_{s})$. The corresponding Fokker-Planck
operator $L_{j}=\tau_{s}^{-1}\left[  \frac{1}{\beta\tilde{\omega}^{2}}%
\frac{\partial^{2}}{\partial q^{2}}+\left(  q-d_{j}\right)  \frac{\partial
}{\partial q}+1\right]  $ describes the diffusion in the effective parabolic
potential
\begin{equation}
U_{j}\left(  q\right)  =E_{j}+\frac{1}{2}\tilde{\omega}^{2}\left(
q-d_{j}\right)  ^{2} \label{eq:U_j}%
\end{equation}
of electronic state $j$ where $\tau_{s}^{-1}=\tilde{D}_{n}\beta\tilde{\omega
}^{2}$ and $\tilde{D}$ is the diffusion coefficient. Going to a dimensionless
generalized coordinate $x=q\tilde{\omega}\sqrt{\beta}$, one can obtain the
equations for the elements of the density matrix $\rho_{ij}(x,t)$ by the
generalization of the equations of Ref.\cite{Fai02JCP_2}. Switching to the
system that rotates with instantaneous frequency%

\begin{align}
\tilde{\rho}_{12}(x,t)  &  =\rho_{12}(x,t)\exp[-i(\omega_{1}t-\varphi
_{1}(t))],\text{ }\tilde{\rho}_{23}(x,t)=\rho_{23}(x,t)\exp[-i(\omega
_{2}t-\varphi_{2}(t))],\nonumber\\
\tilde{\rho}_{13}(x,t)  &  =\rho_{13}(x,t)\exp\{-i[(\omega_{1}+\omega
_{2})t-(\varphi_{1}(t)+\varphi_{2}(t))],
\end{align}
we get%

\begin{align}
\frac{\partial}{\partial t}\rho_{11}(x,t)  &  =\operatorname{Im}[\Omega
_{1}\tilde{\rho}_{12}(x,t)]+L_{1}\rho_{11}(x,t)\nonumber\\
\frac{\partial}{\partial t}\rho_{22}(x,t)  &  =-\operatorname{Im}[\Omega
_{1}\tilde{\rho}_{12}(x,t)+\Omega_{2}^{\ast}\tilde{\rho}_{32}(x,t)]+L_{2}%
\rho_{22}(x,t)+2\Gamma_{32}\rho_{33}(x,t)\nonumber\\
\frac{\partial}{\partial t}\rho_{33}(x,t)  &  =-\operatorname{Im}[\Omega
_{2}\tilde{\rho}_{23}(x,t)]+(L_{3}-2\Gamma_{32})\rho_{33}%
(x,t)\label{eq:diag_matr_IR}\\
\frac{\partial}{\partial t}\tilde{\rho}_{12}(x,t)  &  =i\left[  \omega
_{21}-\omega_{1}(t)-(\hbar\beta)^{-1}x_{2}x\right]  \tilde{\rho}%
_{12}(x,t)+\frac{i}{2}\Omega_{1}^{\ast}[\rho_{22}(x,t)-\rho_{11}%
(x,t)]-\nonumber\\
&  -\frac{i}{2}\Omega_{2}\tilde{\rho}_{13}(x,t)+L_{12}\tilde{\rho}%
_{12}(x,t)\label{eq:rho12tilda}\\
\frac{\partial}{\partial t}\tilde{\rho}_{13}(x,t)  &  =i\left[  \omega
_{31}-\omega_{1}(t)-\omega_{2}(t)-(\hbar\beta)^{-1}x_{3}x\right]  \tilde{\rho
}_{13}(x,t)+\frac{i}{2}\Omega_{1}^{\ast}\tilde{\rho}_{23}(x,t)-\nonumber\\
&  -\frac{i}{2}\Omega_{2}^{\ast}\tilde{\rho}_{12}(x,t)+(L_{13}-\Gamma
_{32})\tilde{\rho}_{13}(x,t)\label{eq:nondiag_matr_IR}\\
\frac{\partial}{\partial t}\tilde{\rho}_{23}(x,t)  &  =i\left[  (\omega
_{31}-\omega_{21})-\omega_{2}(t)-(\hbar\beta)^{-1}\left(  x_{3}-x_{2}\right)
x\right]  \tilde{\rho}_{23}(x,t)+\frac{i}{2}\Omega_{2}^{\ast}\left(  t\right)
[\rho_{33}(x,t)-\rho_{22}(x,t)]+\nonumber\\
&  +\frac{i}{2}\Omega_{1}\tilde{\rho}_{13}(x,t)+(L_{23}-\Gamma_{32}%
)\tilde{\rho}_{23}(x,t) \label{eq:rho23tilda}%
\end{align}
where $\Omega_{1}=D_{21}\mathcal{E}_{1}/\hbar$ and $\Omega_{2}=D_{32}%
\mathcal{E}_{2}/\hbar$ are the Rabi frequencies for transitions $1\rightarrow
2$ and $2\rightarrow3$, respectively. Here $\omega_{i1}=\omega_{i1}^{el}%
+x_{i}^{2}/(2\hbar\beta)$ is the frequency of Franck-Condon transition
$1\rightarrow i,$ $\omega_{ij}^{el}=(E_{i}-E_{j})/\hbar$ is the frequency of
purely electronic transition $j\rightarrow i$, $D_{ij}$ are matrix elements of
the dipole moment operator, $2\Gamma_{32}$ is a probability of nonradiative
transition $3\rightarrow2$ for the excited state absorption problem (see
below); $|x_{j}|=(\hbar\beta\omega_{st}^{1j})^{1/2}$ is a dimensionless shift
between the potential surfaces of states $1$ and $j$ ($x_{1}=0$), which is
related to the corresponding Stokes shift $\omega_{st}^{1j}$ of the
equilibrium absorption and luminescence spectra for transition $1\rightarrow
j$. The last magnitude can be written as$\ \omega_{st}^{1j}=\hbar\beta
\sigma_{2s}^{1j}$ where $\sigma_{2s}^{1j}$ denotes the LF vibration
contribution to a second central moment of an absorption spectrum for
transition $1\rightarrow j$. The terms%

\begin{equation}
L_{j}=\tau_{s}^{-1}\left(  \frac{\partial^{2}}{\partial x^{2}}+\left(
x-x_{j}\right)  \frac{\partial}{\partial x}+1\right)  \label{eq:Ljj}%
\end{equation}
on the right-hand side of Eqs.(\ref{eq:diag_matr_IR}) describe the diffusion
in the corresponding effective parabolic potential
\begin{equation}
U_{j}(x)=E_{j}+\frac{1}{2\beta}(x-x_{j})^{2}\text{ \ \ \ }(j=1,2,3),
\label{eq:Uj}%
\end{equation}
$L_{ij}=(L_{i}+L_{j})/2.$

The partial density matrix of the system $\tilde{\rho}_{ij}\left(  x,t\right)
$ describes the system distribution with a given value of $x$ at time $t$. The
complete density matrix averaged over the stochastic process which modulates
the system energy levels, is obtained by integration of $\tilde{\rho}%
_{ij}\left(  x,t\right)  $ over the generalized coordinate $x$:
\begin{equation}
\langle\tilde{\rho}\rangle_{ij}\left(  t\right)  =\int\tilde{\rho}_{ij}\left(
x,t\right)  dx \label{eq:rhoav}%
\end{equation}
where diagonal quantities $\langle\rho\rangle_{jj}\left(  t\right)  $ are
nothing more nor less than the populations of the electronic states:
$\langle\rho\rangle_{jj}\left(  t\right)  \equiv n_{j}$, $n_{1}+n_{2}+n_{3}=1
$.

We solve coupled Eqs.(\ref{eq:diag_matr_IR})-(\ref{eq:rho23tilda}), using a
basis set expansion with eigenfunctions of diffusion operator $L_{13}$,
similar to Ref. \cite{Fai02JCP_2}.

The solutions, corresponding to the procedure described in this section, are
termed the total model for short, bearing in mind that they take into account
all the relaxations (diffusions) related to populations and electronic
coherences between all the electronic states.

\section{Approximate models}

\label{sec:approx_models}

In this section we describe a number of approaches to the total model
(Eqs.(\ref{eq:diag_matr_IR})-(\ref{eq:rho23tilda})).

\subsection{System with frozen nuclear motion}

\label{subsubsec:no_relaxation}

For pulses much shorter than $\tau_{s}$ one can ignore all the terms $\sim
L_{i},L_{ij}$ on the right-hand sides of Eqs.(\ref{eq:diag_matr_IR}%
)-(\ref{eq:rho23tilda}). It means that our system can be described as an
ensemble of independent three-level systems with different transition
frequencies corresponding to a pure inhomogeneously broadened electronic
transitions. In this case the density matrix equations can be integrated
independently for each $x$. After this the result must be averaged over $x$.
Solutions of the undamped equations for the density matrix are interesting
from the point of view of evaluation of the greatest possible population of
excited states due to coherent effects, because these solutions ignore all the
irreversible relaxations destructing coherence. In addition, a comparison
between the latter solutions and calculations for the total model enables us
to clarify the role of relaxation in the chirp dependence of population
transfer (see Sec.\ref{sec:adiabatic_pop_transfer} below). The approach under
discussion in this section is termed \textquotedblright
relaxation-free\textquotedblright\ model for short.

\subsection{Semiclassical (Lax) approximation}

\label{subsec:Lax}

For broad electronic transitions satisfying the \textquotedblright slow
modulation\textquotedblright\ limit, we have $\sigma_{2s}^{ij}\tau_{s}^{2}%
\gg1$, where $\sigma_{2s}^{ij}$ is the LF vibration contribution to a second
central moment of an absorption spectrum for transition $i\rightarrow j$. In
the last case electronic dephasing is fast, and one can use a semiclassical
(short time) approximation \cite{Lax52}. This limit is also known as the case
of appreciable Stokes losses because the perturbation of the nuclear system
under electronic excitation $i\rightarrow j$ (a quantity $W_{j}-W_{i}$) is
large. Then one can ignore the last term $L_{ij}\tilde{\rho}_{ij}(x,t)$ on the
right-hand side of the corresponding equation for the nondiagonal element of
the density matrix \cite{Fai02JCP_2,Fai90CP,Fai98,Fai04CP} that describes
relaxation (diffusion) of $\tilde{\rho}_{ij}(x,t)$ (Eqs.(\ref{eq:rho12tilda})
and (\ref{eq:rho23tilda})). The solutions, which correspond to disregarding
terms $L_{ij}\tilde{\rho}_{ij}(x,t)$ for broad electronic transitions
$i\rightarrow j$ are termed \textquotedblright partial
relaxation\textquotedblright\ model for short \cite{Fai02JCP_2}. It is worthy
to note that the \textquotedblright partial relaxation\textquotedblright%
\ model offers a particular advantage over the total model. The point is that
the first can be derived not assuming the standard adiabatic elimination of
the momentum $p$ for the non-diagonal density matrix \cite{Fai04CP}, which is
incorrect in the \textquotedblright slow modulation\textquotedblright\ limit
\cite{Pol03}. This issue is quite important in the light of the limits imposed
on Eqs.(\ref{eq:rho12tilda}) and (\ref{eq:rho23tilda}) for nondiagonal
elements of the density matrix \cite{Fra99,Goy01}.

Indeed, in the Wigner representation \cite{Wig32,Hil84,Muk95} equation for
$\tilde{\rho}_{12}$ may be written in the rotating frame as (see
Eq.(\ref{eq:rho12tilda}))%

\begin{align}
\frac{\partial}{\partial t}\tilde{\rho}_{W12}(q,p,t)  &  =i[(U_{2}\left(
q\right)  -U_{1}\left(  q\right)  )/\hbar-\omega_{1}(t)]\tilde{\rho}%
_{W12}(q,p,t)-\frac{i}{2}\Omega_{2}\tilde{\rho}_{W13}(q,p,t)+\nonumber\\
&  +\frac{i}{2}\Omega_{1}^{\ast}[\rho_{W22}(q,p,t)-\rho_{W11}(q,p,t)]+L_{FP12}%
\tilde{\rho}_{W12}(q,p,t) \label{eq:vibroncohW}%
\end{align}
Eq.(\ref{eq:vibroncohW}) has been derived for harmonic potentials,
Eq.(\ref{eq:U_j}), by generalization of equations of
Refs.\cite{Gar85,Har00,Pol03,Fai04CP} where%

\[
L_{FP12}=-p\frac{\partial}{\partial q}+\frac{\partial}{\partial p}\left[
\frac{\gamma}{\beta}\frac{\partial}{\partial p}+\gamma p+\frac{1}{2}\frac
{d}{dq}\left(  U_{1}\left(  q\right)  +U_{2}\left(  q\right)  \right)
\right]
\]
is the Fokker-Planck operator for overdamped Brownian oscillator with
attenuation constant $\gamma$.

In the case of appreciable Stokes losses when the perturbation of the nuclear
system under electronic excitation $1\rightarrow2$ (a quantity $(U_{2}\left(
q\right)  -U_{1}\left(  q\right)  )/\hbar-\omega_{21}^{el}$) is large, the
quantity $\tilde{\rho}_{W12}(q,p,t)$ oscillates fast due to the first term on
the right-hand side of Eq.(\ref{eq:vibroncohW}) (see also Ref.\cite{Pol03}).
Therefore, to the first approximation, on can neglect changes of $\tilde{\rho
}_{W12}(q,p,t)$ due to the last term on the right-hand side of
Eq.(\ref{eq:vibroncohW}). Neglecting this term, integrating both side of
Eq.(\ref{eq:vibroncohW}) over momentum, and bearing in mind that%

\begin{equation}
\tilde{\rho}_{ij}(q,t)=\int_{-\infty}^{\infty}\tilde{\rho}_{Wij}(q,p,t)dp
\end{equation}
and $x=q\tilde{\omega}\sqrt{\beta}$, we get%

\begin{equation}
\frac{\partial}{\partial t}\tilde{\rho}_{12}(x,t)=i[\omega_{21}-\omega
_{1}(t)-(\hbar\beta)^{-1}x_{2}x]\tilde{\rho}_{12}(x,t)+\frac{i}{2}\Omega
_{1}^{\ast}[\rho_{22}(x,t)-\rho_{11}(x,t)]-\frac{i}{2}\Omega_{2}\tilde{\rho
}_{13}(x,t) \label{eq:rho12tilda_without}%
\end{equation}
that is nothing more nor less Eq.(\ref{eq:rho12tilda}) without the last term
$L_{12}\tilde{\rho}_{12}(x,t)$ on the right-hand side. As a matter of fact, a
derivation of Eq.(\ref{eq:rho12tilda_without}) does not involve the assumption
that the momentum is instantly equilibrated. The same can be done with
Eq.(\ref{eq:rho23tilda}) for $\tilde{\rho}_{23}$.

\section{Adiabatic population transfer in the presence of excited-state
absorption}

\label{sec:adiabatic_pop_transfer}

\begin{figure}
[ptb]
\begin{center}
\includegraphics[
height=2.2609in,
width=3.4255in
]%
{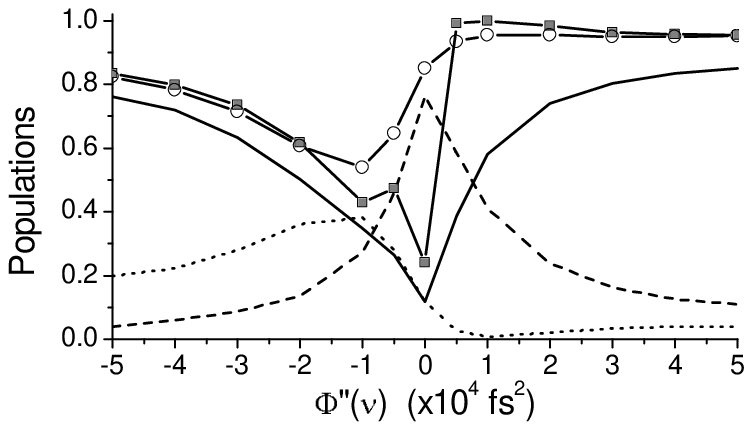}%
\caption{Populations of electronic states after the completion of the pulse
action as functions of $\Phi^{\prime\prime}(\nu)$ in a three-state system.
Calculations without decay of the upper state $3$ into state $2$: $n_{1}$
(dotted line), $n_{2}$ (solid line), $n_{3}$ (dashed line). Line with hollow
circles - $n_{2}$ in the model with fast decay $3\rightarrow2$ $\ \Gamma_{32}%
$=10 ps$^{-1}$. For comparison we also show $n_{2}$ for a two-state system
(line with squares). Total relaxation model with diffusion of all matrix
elements.}%
\label{fig:ESA}%
\end{center}
\end{figure}

We shall study the ESA effects on ARP in complex molecules by the example of
Coumarin 153 in liquid solution \cite{Kov97}. In the frequency domain, the
electric field can be written as $|E(\tilde{\omega})|\exp[i\Phi(\tilde{\omega
})]$ and the phase term $\Phi(\tilde{\omega})$ can be expanded in a Taylor
series $\Phi(\tilde{\omega})=\Phi(\omega)+(1/2)\Phi^{\prime\prime}%
(\omega)(\tilde{\omega}-\omega)^{2}+...$ We shall consider linear chirped
pulses of the form%

\begin{equation}
E(t)=\mathcal{E}_{0}\exp[-\frac{1}{2}(\delta^{2}-i\mu)(t-t_{0})^{2}]
\label{eq:gausspulse}%
\end{equation}
where the parameters $\delta$ and $\mu$ are determined by the formulae
\cite{Cer96,Fai98}:%

\begin{equation}
\delta^{2}=2\{\tau_{p0}^{2}+\left[  2\Phi^{\prime\prime}\left(  \omega\right)
/\tau_{p0}\right]  ^{2}\}^{-1},\text{ }\mu=-4\Phi^{\prime\prime}\left(
\omega\right)  \left[  \tau_{p0}^{4}+4\Phi^{\prime\prime2}\left(
\omega\right)  \right]  ^{-1}, \label{eq:deltamu}%
\end{equation}
$\tau_{p0}=t_{p0}/\sqrt{2\ln2}$, $t_{p0}$ is the pulse duration of the
corresponding transform-limited pulse. Fig.\ref{fig:ESA} shows populations of
electronic states after the completion of the one pulse action as functions of
the chirp rate in the frequency domain $\Phi^{\prime\prime}(\nu)=4\pi^{2}%
\Phi^{\prime\prime}\left(  \omega\right)  $. For the molecule under
consideration a two-photon resonance occurs at the doubled frequency of the
Franck-Condon transition $1\rightarrow2$. Absorption spectrum corresponding to
transition $1\rightarrow3$ is rather narrow that means $x_{3}=0$.
The values of parameters for Fig.\ref{fig:ESA} were as follows: the pulse
duration of the transform-limited (non-chirped) pulse $t_{p0}=10$ $fs$,
$\omega_{st}^{12}=2686$ $cm^{-1}$, $D_{12}=D_{32}=6$ $D$ \cite{Kov97},
$\tau_{s}=70$ $fs$, the saturation parameter, which is proportional to the
pulse energy \cite{Fai02JCP_2}, $Q^{\prime}\equiv\sqrt{\pi}|D_{12}%
\mathcal{E}_{\max}|^{2}t_{p}/(2\hbar^{2}\sqrt{2\sigma_{2s}^{12}})=5$; the
one-photon resonance for Franck-Condon transition $1\rightarrow2$ occurs at
the pulse maximum, i.e. $\omega=\omega_{21}$.

\begin{figure}
[ptb]
\begin{center}
\includegraphics[
height=2.1926in,
width=3.398in
]%
{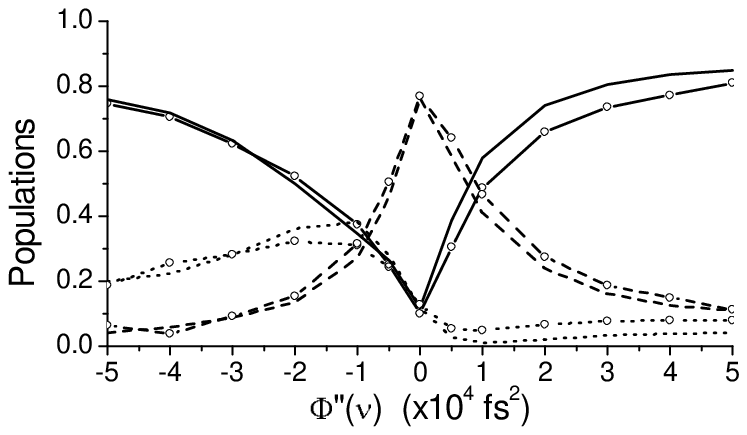}%
\caption{Populations of electronic states $n_{1}$\ (dotted\ lines),\ $n_{2}%
$\ (solid\ lines) and\ $n_{3}$\ (dashed\ lines) after the completion of the
pulse action calculated without decay of the upper state $3$ into state $2$ as
functions of $\Phi^{\prime\prime}(\nu)$. The partial\ relaxation and the total
\ models - lines with and without hollow\ circles, respectively. All the
parameters are identical to those of \ Fig.\ref{fig:ESA}. \ }%
\label{fig:ESA_models}%
\end{center}
\end{figure}

Fig.\ref{fig:ESA_models} contrasts calculations using the total model
(Fig.\ref{fig:ESA}) with those of the partial relaxation model. The latter
includes both diffusion of all the diagonal elements of the density matrix and
one off-diagonal element $\rho_{13}$. The point is that transition
$1\rightarrow3$ occurs without changing the state of vibrational subsystems of
a molecule and a solvent, and therefore can not be described in a
semiclassical (short time) approximation. Fig.\ref{fig:ESA_models} shows a
good agreement between calculation results for the models under consideration.%

One can see from Fig.\ref{fig:ESA}, first, that population $n_{2}$ for a
molecule with a fast decay $3\rightarrow2$, which closely resembles
experimental data \cite{Cer96} for LD690\footnote{According to
Ref.\cite{Sension06JPC}, LD690 shows ESA.}, is distinctly different from that
of a two-state system for $\left\vert \Phi^{\prime\prime}(\nu)\right\vert
<15\cdot10^{3}$ $fs^{2}$ when the excited pulse is rather short. This means
that the excited state absorption has a profound effect on coherent population
transfer in complex molecules. Second, $n_{3}$ strongly decreases when
$|\Phi^{\prime\prime}(\nu)|$ increases.

\begin{figure}
[ptb]
\begin{center}
\includegraphics[
height=2.6184in,
width=4.1901in
]%
{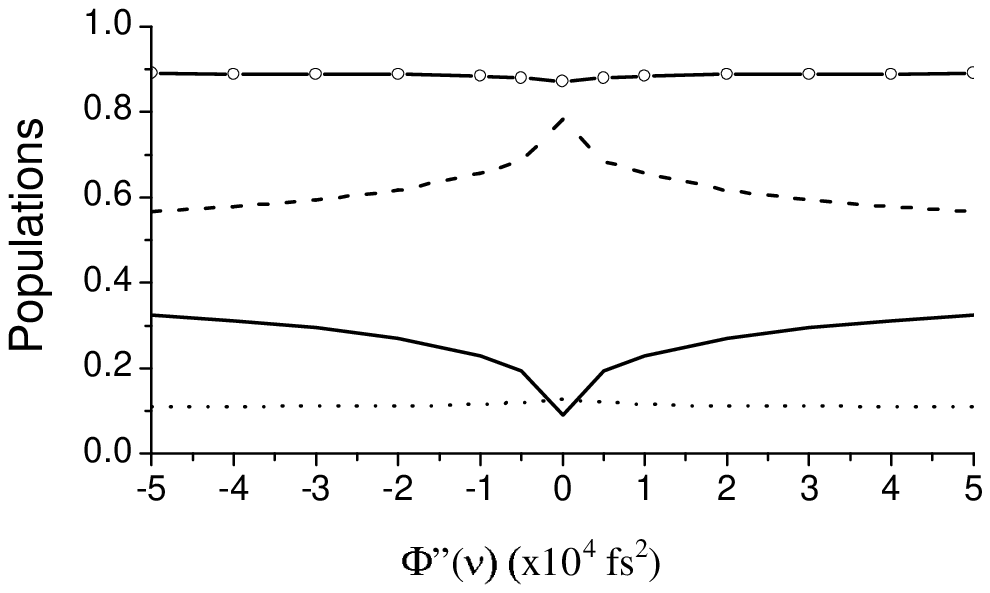}%
\caption{Populations\ of\ electronic\ states $n_{1}$\ (dotted\ line),\ $n_{2}%
$\ (solid\ line),\ $n_{3}$\ (dashed\ line)\ and $n_{2}+n_{3}$ (line with
hollow circles) after the completion of the pulse action as
functions\ of\ $\Phi^{\prime\prime}(\nu)$\ for the\ relaxation-free
model\ $\tau_{s}\rightarrow\infty$. Other parameters are identical to those of
\ Fig.\ref{fig:ESA}. In\ the\ case under
consideration\ the\ combined\ population\ $n_{2}+n_{3}$\ does
not\ depend\ on\ $\Phi^{\prime\prime}(\nu)$.\ \ }%
\label{fig:no_relaxation}%
\end{center}
\end{figure}

To understand these results, we will consider first two transitions
separately. One can obtain the following criterion for the adiabaticity of one
transition in the absence of relaxation: $Q^{\prime}>>1$ where $Q^{\prime}$ is
the saturation parameter. It conforms to the value of $Q^{\prime}=5$ used in
our calculations. The condition $Q^{\prime}>>1$ follows from the adiabatic
criterion for a two-level system:%

\begin{equation}
\left\vert \frac{d\omega(t)}{dt}\right\vert \ll|\Omega_{1,2}(t)|^{2}
\label{eq:ARPf}%
\end{equation}
where $\Omega_{1,2}(t)=|D_{21,32}\mathcal{E}(t)|/\hbar$ are the Rabi
frequencies for transitions $1\rightarrow2$ and $2\rightarrow3$, respectively.
Adiabatic criterion Eq.(\ref{eq:ARPf}) was fulfilled in our simulations for
both transitions $1\rightarrow2$ and $2\rightarrow3$ at any $\Phi
^{\prime\prime}(\nu)$. However, Fig.\ref{fig:ESA} shows that $n_{3}$ strongly
decreases when $|\Phi^{\prime\prime}(\nu)|$ increases. To clarify the reasons
for strong decreasing $n_{3}$ it is instructive to carry out the corresponding
calculations for the relaxation-free model of
Sec.\ref{subsubsec:no_relaxation} shown in Fig.\ref{fig:no_relaxation}.
In\ this\ case excitation\ of\ state\ $3$%
\ with\ a\ transform-limited\ pulse\ is
slightly\ more\ effective\ as\ compared\ to\ a\ strongly\ chirped\ pulse
of\ the\ same\ energy. The point is that a two-photon resonance occurs for a
number of spectral components of a transform-limited pulse and only at the
maximum of a strongly chirped pulse. \
However, Fig.\ref{fig:no_relaxation} does not show strong decreasing the
population of state $3$ when $\Phi^{\prime\prime}(\nu)$ increases. This means
that \textit{relaxation is responsible for strong decreasing} $n_{3}$ as a
function of $\Phi^{\prime\prime}(\nu)$ in spite of the fact that relaxation
does not destroy ARP when the Rabi frequencies exceed the reciprocal
irreversible dephasing time $(T^{\prime})^{-1}$ \cite{Fai04JCP}%

\begin{equation}
\Omega_{1,2}>>1/T^{\prime} \label{eq:Omega>1/T'}%
\end{equation}
The last condition was fulfilled in our simulations at least for
$|\Phi^{\prime\prime}(\nu)|\lesssim10^{4}$ $fs^{2}$.

To clarify this issue, we shall consider a population transfer between
randomly fluctuating levels.

\subsection{Population transfer between randomly fluctuating levels}

\label{subsec:random}

The picture of randomly fluctuating levels \cite{Fai04JCP} offers a simple and
physically clear explanation of numerical results \cite{Fai02JCP_2} obtained
for population transfer in a two-state system. Here we shall generalize the
Landau-Zener (LZ) calculations putting in a third level \cite{Car86_2} to
random crossing of levels.

Let us write the Schr\"{o}dinger equations for the amplitudes of states
$a_{1,2,3}$ for the system under consideration. Switching to new variables
$\tilde{a}_{k}$:%
\begin{equation}
a_{k}=\tilde{a}_{k}\exp\left(  -\frac{i}{\hbar}\int_{0}^{t}U_{2}dt\right)
\text{,} \label{eq:a_k}%
\end{equation}
we obtain in the rotating wave approximation%
\begin{equation}
i\frac{d}{dt}\left(
\begin{array}
[c]{c}%
\tilde{a}_{1}\\
\tilde{a}_{2}\\
\tilde{a}_{3}%
\end{array}
\right)  =\left(
\begin{array}
[c]{ccc}%
(U_{1}-U_{2})/\hbar+\omega_{1}(t) & -\Omega_{1}/2 & 0\\
-\Omega_{1}/2 & 0 & -\Omega_{2}/2\\
0 & -\Omega_{2}/2 & (U_{3}-U_{2})/\hbar-\omega_{2}(t)
\end{array}
\right)  \left(
\begin{array}
[c]{c}%
\tilde{a}_{1}\\
\tilde{a}_{2}\\
\tilde{a}_{3}%
\end{array}
\right)  \label{eq:Schr2}%
\end{equation}
Throughout this section effective parabolic potentials (\ref{eq:Uj}) are
considered as functions of generalized coordinate $\alpha=x\sqrt{\sigma
_{2s}^{12}}-\omega_{st}^{12}$: $U_{j}(\alpha)=E_{j}+\frac{\hbar}{2\omega
_{st}^{12}}\{\alpha+\sqrt{\omega_{st}^{12}}[\sqrt{\omega_{st}^{12}%
}+(-1)^{sgn(x_{j})}\sqrt{\omega_{st}^{1j}}]\}^{2}$. Here
\begin{equation}
(U_{3}-U_{2})/\hbar-\omega_{2}(t)=[(\omega_{32}^{el}+\omega_{st}%
^{12}/2)-\omega_{2}]+\alpha+\mu_{2}t \label{eq:U3-2}%
\end{equation}
for $x_{3}=x_{1}$ (that corresponds to Coumarin 153), and
\begin{equation}
(U_{1}-U_{2})/\hbar+\omega_{1}(t)=[\omega_{1}-(\omega_{21}^{el}-\omega
_{st}^{12}/2)]+\alpha-\mu_{1}t, \label{eq:U1-2}%
\end{equation}
for linear chirped pulses $\omega_{1,2}\left(  t\right)  =\omega_{1,2}%
-\mu_{1,2}t$.

Let us define instantaneous crossings of state $2$ with photonic repetitions
$1^{\prime}$ and $3^{\prime}$ of states $1$ and $3$, respectively. They are
determined by the conditions that quantities Eqs.(\ref{eq:U3-2}) and
(\ref{eq:U1-2}) are equal to zero:%

\begin{align}
\alpha_{12}(t)  &  =(\omega_{21}^{el}-\omega_{st}^{12}/2)-\omega_{1}+\mu
_{1}t\equiv\alpha_{12}(0)+\mu_{1}t\label{eq:crossing_points}\\
\alpha_{23}(t)  &  =\omega_{2}-(\omega_{32}^{el}+\omega_{st}^{12}/2)-\mu
_{2}t\equiv\alpha_{23}(0)-\mu_{2}t\nonumber
\end{align}
Near the intersection points one can consider $\alpha$ as a linear function of
time. For small $t,$ $\alpha(t)\approx\alpha_{12}(0)+\dot{\alpha}t$. Let
$\alpha_{12}(0)=\alpha_{23}(0)$, i.e. states $2$, $1^{\prime}$ and $3^{\prime
}$ cross at the same point when $t=0$. This means%

\begin{equation}
\omega_{21}^{el}+\omega_{32}^{el}=\omega_{1}+\omega_{2}
\label{eq:two_phot_resonance}%
\end{equation}
i.e. the two-photon resonance occurs for $t=0$. Then Eqs.(\ref{eq:Schr2}) take
the following form%

\begin{equation}
i\frac{d}{dt}\left(
\begin{array}
[c]{c}%
\tilde{a}_{1}\\
\tilde{a}_{2}\\
\tilde{a}_{3}%
\end{array}
\right)  =\left(
\begin{array}
[c]{ccc}%
(\dot{\alpha}-\mu_{1})t & -\Omega_{1}/2 & 0\\
-\Omega_{1}/2 & 0 & -\Omega_{2}/2\\
0 & -\Omega_{2}/2 & (\dot{\alpha}+\mu_{2})t
\end{array}
\right)  \left(
\begin{array}
[c]{c}%
\tilde{a}_{1}\\
\tilde{a}_{2}\\
\tilde{a}_{3}%
\end{array}
\right)  \label{eq:Schr3}%
\end{equation}
that can be reduced to Eqs.(2) of Ref.\cite{Car86_2}. Using the solution
obtained in \cite{Car86_2} and considering identical chirps when $\mu_{1}%
=\mu_{2}\equiv\mu$, we get for the initial condition $\left\vert a_{1}%
(-\infty)\right\vert ^{2}=1,$ $\left\vert a_{2,3}(-\infty)\right\vert ^{2}=0$%
\begin{equation}
\left\vert a_{3}(\infty)\right\vert ^{2}=\left\{
\begin{array}
[c]{c}%
(1-P)(1-Q)\text{ for }-|\mu|<\dot{\alpha}<|\mu|\\
P(1-P)(1-Q)\text{ for both }\dot{\alpha}>-\mu\text{ when }\mu<0\text{, and
}\dot{\alpha}<-\mu\text{ when }\mu>0\\
Q(1-P)(1-Q)\text{ for both }\dot{\alpha}<\mu\text{ when }\mu<0\text{, and
}\dot{\alpha}>\mu\text{ when }\mu>0
\end{array}
\right\}  \label{eq:n3}%
\end{equation}
where%

\begin{equation}
P=\exp\left(  -\frac{\pi\Omega_{1}^{2}}{4|\dot{\alpha}-\mu|}\right)  \text{,
}Q=\exp\left(  -\frac{\pi\Omega_{2}^{2}}{4|\dot{\alpha}+\mu|}\right)
\label{eq:PQ_1}%
\end{equation}

Similar to Ref.\cite{Fai04JCP}, we consider $\alpha$ as a stochastic Gaussian
variable. Consequently, we must average Eqs.(\ref{eq:n3}) over random crossing
of levels described by Gaussian random noise induced by intra- and
intermolecular fluctuations. It can be easily done for a
\textit{differentiable (}non-Markovian) Gaussian process \cite{Fai04JCP},
bearing in mind an independence of $\alpha$ and $\dot{\alpha}$ from each other
for such processes. Therefore, we shall consider in this section a
\textit{differentiable (}non-Markovian) Gaussian noise, as opposed to
previuous sections. In addition, we consider a slow modulation limit when
$\sigma_{2s}^{12}\tau_{s}^{2}>>1$. Averaging Eqs.(\ref{eq:n3}), we obtain the
following expression for the population of state $3$ when $\mu>0$%

\begin{align}
n_{3}  &  =\int_{-\infty}^{\infty}d\alpha\lbrack\int_{-\infty}^{-|\mu
|}P(1-P)(1-Q)f(\alpha,\dot{\alpha})d\dot{\alpha}+\int_{-|\mu|}^{|\mu
|}(1-P)(1-Q)f(\alpha,\dot{\alpha})d\dot{\alpha}\nonumber\\
&  +\int_{|\mu|}^{\infty}Q(1-P)(1-Q)f(\alpha,\dot{\alpha})d\dot{\alpha}]
\label{eq:n3_mu+}%
\end{align}
Here $f(\alpha,\dot{\alpha})$ is the joint probability density for $\alpha$
and its derivative $\dot{\alpha}$:%

\begin{equation}
f(\alpha,\dot{\alpha})=\frac{1}{2\pi\sqrt{\sigma_{2s}^{12}(-\ddot{k}(0))}}%
\exp\left[  -\frac{\alpha^{2}}{2\sigma_{2s}}+\frac{\dot{\alpha}^{2}}{2\ddot
{k}(0)}\right]  ,
\end{equation}
$\ddot{k}(0)$ is the second derivative of the correlation function
$k(t)=<\alpha(0)\alpha(t)>=\sigma_{2s}^{12}\exp(-|t|/\tau_{s})$ of the
energetic fluctuations evaluated at zero. Eq.(\ref{eq:n3_mu+}) is written for
$\mu>0$ (negatively chirped pulse). One can easily show that $n_{3}$ is
symmetrical with respect to the chirp sign. The point is that a simple
stochastic model of this Section misses any chromophore's effects on bath, in
particular the dynamical Stokes shift (see Ref.\cite{Fai97CP} for details).
This is opposite to the models of previous sections, which do describe the
dynamical Stokes by the drift term (the second term on the right-hand side of
Eq.(\ref{eq:Ljj})).

Integrating Eq.(\ref{eq:n3_mu+}) with respect to $\alpha$ and entering a
dimensionless variable $y=\dot{\alpha}/\left|  \mu\right|  ,$ we get%

\begin{align}
n_{3}  &  =\sqrt{\frac{\xi}{2\pi}}[\int_{-\infty}^{-1}P(1-P)(1-Q)\exp\left(
-\frac{\xi}{2}y^{2}\right)  dy+\int_{-1}^{1}(1-P)(1-Q)]\exp\left(  -\frac{\xi
}{2}y^{2}\right)  dy\nonumber\\
&  +\int_{1}^{\infty}Q(1-P)(1-Q)\exp\left(  -\frac{\xi}{2}y^{2}\right)  dy]
\label{eq:n3_mu+_1}%
\end{align}
where%
\begin{equation}
P=\exp\left(  -\frac{\Omega_{1}^{2}}{\Omega_{2}^{2}}\frac{\varkappa}%
{2|y-1|}\right)  \text{ and }Q=\exp\left(  -\frac{\varkappa}{2|y+1|}\right)
\text{,} \label{eq:PQ_2}%
\end{equation}
and%

\begin{equation}
\varkappa=\frac{\pi\Omega_{2}^{2}}{2\left\vert \mu\right\vert }>0,\text{ }%
\xi=-\frac{\mu^{2}}{\ddot{k}(0)}>0 \label{eq:DP}%
\end{equation}
are dimensionless parameters.

When adiabatic criterion Eq.(\ref{eq:ARPf}) is satisfied, parameter
$\varkappa$ is much larger than $1$ since $\left\vert d\omega(t)/dt\right\vert
=|\mu|$ for a linear chirped pulse. Then the integrals on the right-hand side
of Eq.(\ref{eq:n3_mu+_1}) can be evaluated by the method of Laplace, similar
to Ref.\cite{Fai04JCP}. The result is especially simple for strong
interaction, Eq.(\ref{eq:Omega>1/T'}), where the irreversible dephasing time
of transitions $1\rightarrow2$ and $2\rightarrow3$ is given by \cite{Fai04JCP}
$T^{\prime}=1/[-\ddot{k}(0)]^{1/4}$. Then, as one can see also from
Eqs.(\ref{eq:n3_mu+_1}) and (\ref{eq:PQ_2}), the main contribution to $n_{3}$
is given by%

\begin{equation}
n_{3}\simeq\sqrt{\frac{\xi}{2\pi}}\int_{-1}^{1}\exp\left(  -\frac{\xi}{2}%
y^{2}\right)  dy=\operatorname{erf}\left(  \frac{|\mu|T^{\prime2}}{\sqrt{2}%
}\right)  \label{eq:n3_mu+_3}%
\end{equation}
Since $\operatorname{erf}(1.5)=0.966$, we obtain that relaxation does not
hinder a population transfer to state $3$ when%

\begin{equation}
|\mu|T^{\prime2}\geq2 \label{eq:counter-movement}%
\end{equation}
For strongly chirped pulses \cite{Fai00CPL}, $\mu|T^{\prime2}/\sqrt{2}%
\approx2\sqrt{2}\pi^{2}T^{\prime2}/|\Phi^{\prime\prime}(\nu)|$.

\begin{figure}
[ptb]
\begin{center}
\includegraphics[
height=2.6521in,
width=4.4952in
]%
{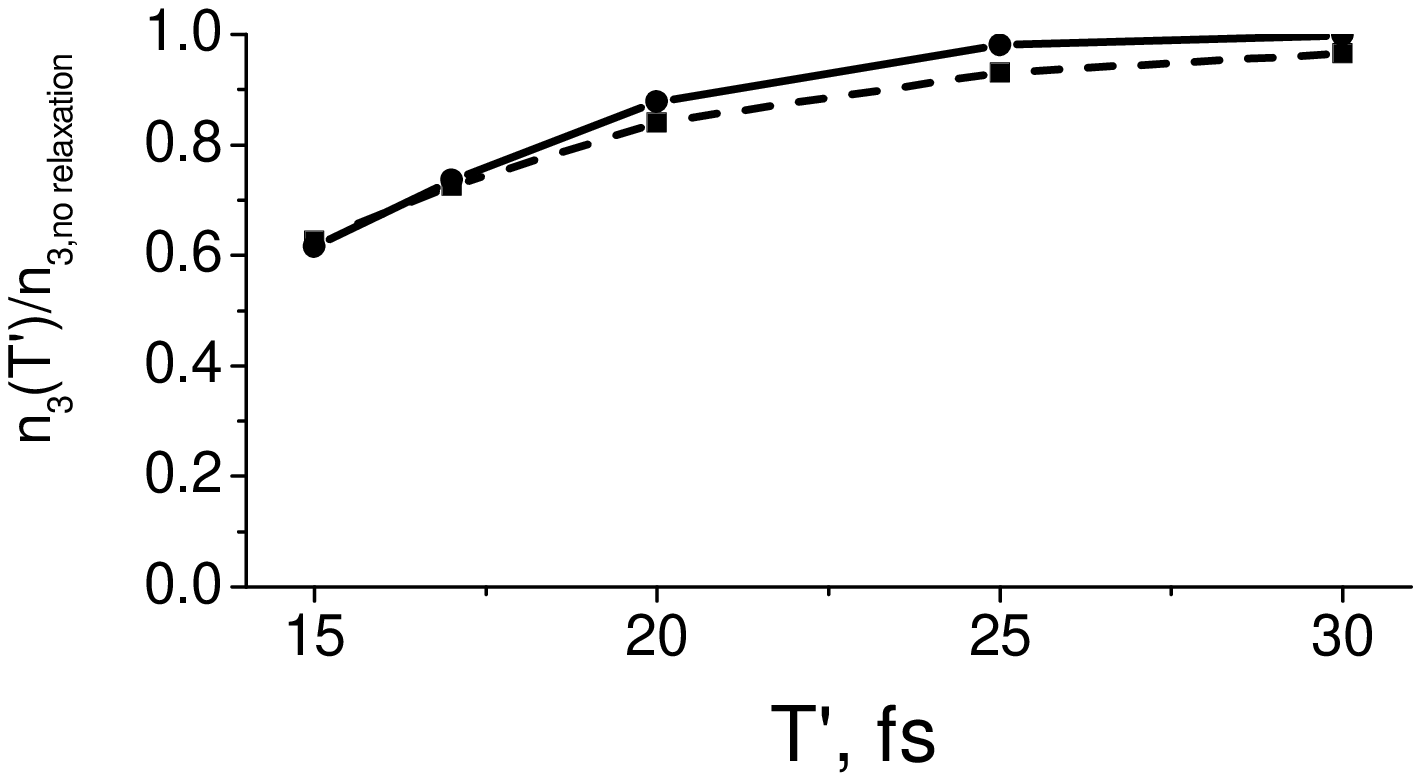}%
\caption{Population of state $3$ as a function of the irreversible dephasing
time $T^{\prime}$ for $\Phi^{\prime\prime}(\nu)=10^{4}$ fs$^{2}$ calculated by
Eq.(\ref{eq:n3_mu+_3}) (solid line with circles) and numerical solution of
Eqs.(\ref{eq:diag_matr_IR})-(\ref{eq:rho23tilda}) (dashed line with squares).
$n_{3,\text{no relaxation}}\equiv n_{3}(T^{\prime}\rightarrow\infty)$. Other
parameters are identical to those of \ Fig.\ref{fig:ESA}.}%
\label{fig:random versus simulation}%
\end{center}
\end{figure}

Eq.(\ref{eq:counter-movement}) expresses an extra criterion for coherent
population transfer to those we have obtained before for a two-level system
\cite{Fai04JCP}, Eqs.(\ref{eq:ARPf}) and (\ref{eq:Omega>1/T'}). New criterion
(\ref{eq:counter-movement}) implies conservation of the \textquotedblleft
counter-movement\textquotedblright\ of the \textquotedblleft photonic
repetitions\textquotedblright\ of states $1$ and $3$, in spite of random
crossing of levels. Condition (\ref{eq:counter-movement}) is exemplified by
Fig.\ref{fig:random versus simulation}. In addition,
Fig.\ref{fig:random versus simulation} shows an excellent agreement of simple
formula (\ref{eq:n3_mu+_3}) with numerical calculations. It is worthy to note
that condition (\ref{eq:Omega>1/T'}) was fulfilled in our simulations, though
in the last case $T^{\prime}=(\tau_{s}/\sigma_{2s})^{1/3}$ is determined
independently of $\ddot{k}(0)$ \cite{Fai90OS}, which does not exist for the
Gaussian-Markovian process.%

\subsection{Influence of excited-state absorption when detuning from
two-photon resonance occurs}

For Coumarin 153 in liquid solution considered above a two-photon resonance
occurs at the doubled frequency of the Franck-Condon transition $1\rightarrow
2$. In this section we consider populations of electronic states when the
condition for two-photon resonance is violated.
\begin{figure}
[ptb]
\begin{center}
\includegraphics[
height=2.0516in,
width=3.1985in
]%
{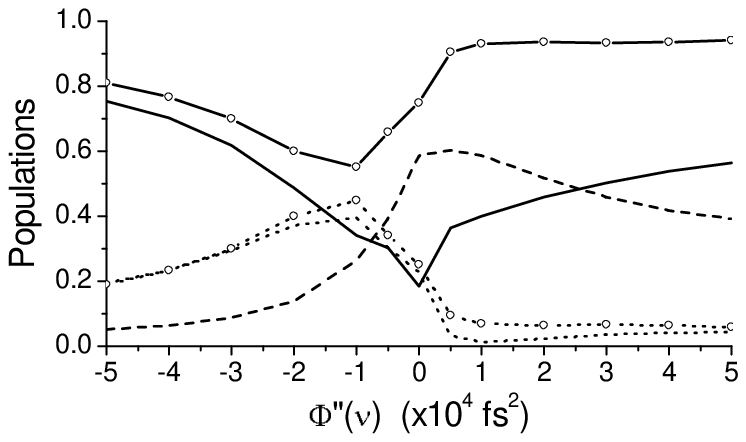}%
\caption{Populations of electronic states after the completion of the pulse
action as functions of $\Phi^{\prime\prime}(\nu)$ in a three-state system. The
frequency of purely electronic transition $3\rightarrow2$, $\omega_{32}^{el}$,
decreases by $\omega_{st}^{12}/4$ with the conservation of $x_{3}=0$.
Calculations without decay of the upper state $3$ into state $2$: $n_{1}$
(dotted line), $n_{2}$ (solid line), $n_{3}$ (dashed line). The corresponding
populations in the model with fast decay $3\rightarrow2$ $\ \Gamma_{32}=10$
$ps^{-1}$ are shown by the same lines with hollow circles. }%
\label{fiq:detuning_down}%
\end{center}
\end{figure}
\begin{figure}
[ptbptb]
\begin{center}
\includegraphics[
height=2.1323in,
width=3.3271in
]%
{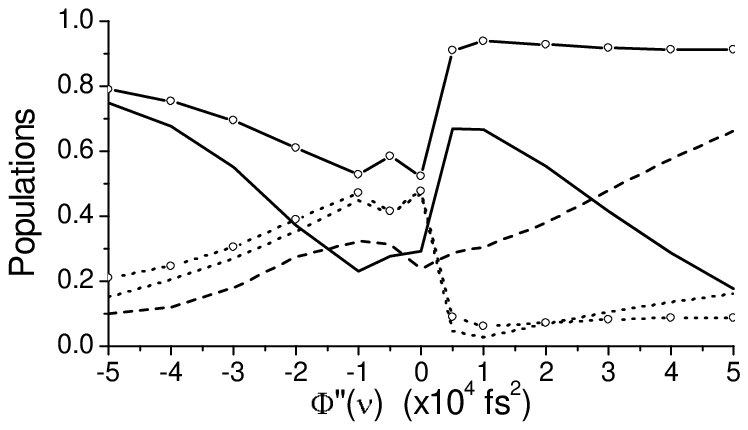}%
\caption{Populations of electronic states after the completion of the pulse
action as functions of $\Phi^{\prime\prime}(\nu)$ in a three-state system.
Equilibrium position of state $3$ is offset to the right by $x_{3}=$ $x_{2}/2$
and down so that frequencies of Franck-Condon transitions $1\rightarrow2$ and
$2\rightarrow3$ are equal: $\omega_{21}=\omega_{32}$. Calculations without
decay of the upper state $3$ into state $2$: $n_{1}$ (dotted line), $n_{2}$
(solid line), $n_{3}$ (dashed line). The corresponding populations in the
model with fast decay $3\rightarrow2$ $\ \Gamma_{32}=10$ $ps^{-1}$ are shown
by the same lines with hollow circles. }%
\label{fiq:detuning_right}%
\end{center}
\end{figure}
Figs. \ref{fiq:detuning_down} and \ref{fiq:detuning_right} show populations of
electronic states for the total model after the completion of the pulse action
as functions of $\Phi^{\prime\prime}(\nu)$ for the same values of parameters
as for Fig.\ref{fig:ESA} with the only difference concerning the position of
state $3$. The frequency of purely electronic transition $3\rightarrow2$
$\omega_{32}^{el}$ decreases by $\omega_{st}^{12}/4$ with the conservation of
$x_{3}=0$ for Fig. \ref{fiq:detuning_down}. Equilibrium position of state $3$
is offset to the right by $x_{3}=$ $x_{2}/2$ and down so that frequencies of
Franck-Condon transitions $1\rightarrow2$ and $2\rightarrow3$ are equal:
$\omega_{21}=\omega_{32}$ for Fig. \ref{fiq:detuning_right}.

One can see from Figs.\ref{fig:ESA}, \ref{fiq:detuning_down} and
\ref{fiq:detuning_right}, first, that population $n_{1}$ and, as a
consequence, $n_{2}+n_{3}$ depend only slightly on the occurrence of fast
decay $3\rightarrow2$. Second, populations $n_{2}$ and $n_{3}$ in the absence
of fast decay $3\rightarrow2$ are very sensitive to the violation of the
two-photon resonance condition. However, a behavior of $n_{2}$, when fast
decay $3\rightarrow2$ occurs, and $n_{1}$ as functions of $\Phi^{\prime\prime
}(\nu)$ is very similar for the figures under discussion, regadless of the
two-photon resonance condition. Experimental measurements commonly correspond
to $n_{2}$ and are carried out under the fast decay $3\rightarrow2$
conditions. Thus the behavior of $n_{2}$ for fast decay $3\rightarrow2$ shown
in Figs.\ref{fig:ESA}, \ref{fiq:detuning_down} and \ref{fiq:detuning_right} is
rather versatile.

\section{Population transfer in the presence of two-exciton processes.
Selective excitation of single and two-exciton states with chirped pulses}

\label{sec:excitons}

Consider a dimer of chromophores each with two electronic states described by
the Frenkel exciton Hamiltonian \cite{Dav71,Fle99,Muk04ChemRev} and excited
with electromagnetic field Eq.(\ref{eq:field}). The Hamiltonian of the dimer
is given by%

\begin{equation}
H=\sum_{m=1,2}\hbar\bar{\Omega}_{m}B_{m}^{+}B_{m}+\hbar J(B_{1}^{+}B_{2}%
+B_{2}^{+}B_{1})+H_{bath}+H_{eb}-\sum_{m=1,2}\mathbf{D}_{m}\cdot
\mathbf{E}(t)(B_{m}^{+}+B_{m}) \label{eq:Hdimer}%
\end{equation}
where $B_{m}^{+}=|m\rangle\langle0|$ $(B_{m}=|0\rangle\langle m|)$ are exciton
creation (annihilation) operators associated with the chromophore $m$, which
satisfy the commutation \ rules $[B_{n},B_{m}^{+}]=\delta_{nm}(1-2B_{m}%
^{+}B_{m}),$ $\delta_{nm}$ is the Kroenecker delta; $|0\rangle$ and
$|m\rangle$ denote the ground state and a state corresponding to the
excitation of chromophore $m$, respectively. $\mathbf{D}_{m}$ is the
transition dipole moment of molecule $m$, $H_{bath}$ represents a bath and
$H_{eb}$ its coupling with the exciton system. We assume that the bath is
harmonic and that the coupling is linear in the nuclear coordinates
\begin{equation}
H_{eb}=-\hbar\sum_{mn}\alpha_{mn}B_{m}^{+}B_{n} \label{eq:Heb}%
\end{equation}
where $\alpha_{mn}$ represent collective bath coordinates. $\hbar\bar{\Omega
}_{1}$($\hbar\bar{\Omega}_{2}$) and $\hbar J$ are the exciton energy of $1$
($2$) chromophore and their coupling energy at the equilibrium nuclear
coordinate of the ground electronic state. One can consider $\alpha_{mn}$ as
diagonal: $\alpha_{mn}=\alpha_{m}\delta_{nm}$ on the assumption that the
electronic coupling constant fluctuation amplitude is negligibly smaller than
the site energy fluctuation amplitude \cite{Fle99}.

Diagonalizing the electronic Hamiltonian%

\begin{equation}
H_{e}=\sum_{m=1,2}\hbar\bar{\Omega}_{m}B_{m}^{+}B_{m}+\hbar J(B_{1}^{+}%
B_{2}+B_{2}^{+}B_{1}) \label{eq:H_e}%
\end{equation}
by unitary transformation \cite{Cho_Fle05}%
\begin{equation}
U^{-1}=\left(
\begin{array}
[c]{cc}%
\cos\theta & \sin\theta\\
-\sin\theta & \cos\theta
\end{array}
\right)  \label{eq:U-1}%
\end{equation}
where%

\begin{equation}
\tan2\theta=\frac{2J}{\bar{\Omega}_{1}-\bar{\Omega}_{2}},\text{ }0<\theta
<\pi/2, \label{eq:tan2teta}%
\end{equation}
one can get the eigenstates for the one-exciton states $|e_{i}\rangle$ and the
transition dipole moments $D_{e_{i}}$ ($i=1,2$) corresponding to the
transitions between the ground and single-exciton states as%

\begin{equation}
\left(
\begin{array}
[c]{c}%
a_{e_{1}}\\
a_{e_{2}}%
\end{array}
\right)  =U^{-1}\left(
\begin{array}
[c]{c}%
A_{1}\\
A_{2}%
\end{array}
\right)  =\left(
\begin{array}
[c]{c}%
A_{1}\cos\theta+A_{2}\sin\theta\\
-A_{1}\sin\theta+A_{2}\cos\theta
\end{array}
\right)  \label{eq:one_exciton}%
\end{equation}
Here $a_{e_{i}}=|e_{i}\rangle,D_{e_{i}}$ and $A_{i}=B_{i}^{+}|0\rangle,D_{i}$;
$D_{1}$ and $D_{2}$ are the site transition moments. The two one-exciton
energies are given by%
\begin{align}
\hbar\bar{\Omega}_{e_{1}}  &  =\hbar\bar{\Omega}_{1}\cos^{2}\theta+\hbar
\bar{\Omega}_{2}\sin^{2}\theta+\hbar J\sin2\theta
,\label{eq:1exciton_eigenvalues}\\
\hbar\bar{\Omega}_{e_{2}}  &  =\hbar\bar{\Omega}_{1}\sin^{2}\theta+\hbar
\bar{\Omega}_{2}\cos^{2}\theta-\hbar J\sin2\theta,\nonumber
\end{align}
The two-exciton state wavefunction and its energy are as following%

\begin{equation}
|e_{3}\rangle=B_{1}^{+}B_{2}^{+}|0\rangle\equiv B_{e_{3}}^{+}|0\rangle
\label{eq:two_exciton}%
\end{equation}

\begin{equation}
\hbar\bar{\Omega}_{e_{3}}=\hbar\bar{\Omega}_{1}+\hbar\bar{\Omega}_{2}
\label{eq:2exciton_eigenvalue}%
\end{equation}
The transition dipole moments between the single-exciton and two-exciton
states are given by%

\begin{equation}
D_{e_{1}e_{3}}=D_{1}\sin\theta+D_{2}\cos\theta,\text{ }D_{e_{2}e_{3}}%
=D_{1}\cos\theta-D_{2}\sin\theta\label{eq:dipole_between1exciton_2exciton}%
\end{equation}
However, the transition between the ground and two-exciton states is not allowed.

In the eigenstate representation, the Hamiltonian of Eq.(\ref{eq:Hdimer}) is
rewritten as%
\begin{align}
H  &  =\sum_{i=1,2,3}\hbar(\bar{\Omega}_{e_{i}}-\alpha_{e_{i}})B_{e_{i}}%
^{+}B_{e_{i}}-\hbar\sum_{\substack{i,j=1,2\\i\neq j}}\alpha_{e_{i}e_{j}%
}B_{e_{i}}^{+}B_{e_{j}}+H_{bath}-\nonumber\\
&  -\sum_{i=1,2}[\mathbf{D}_{e_{i}}(B_{e_{i}}^{+}+B_{e_{i}})+\mathbf{D}%
_{e_{i}e_{3}}(B_{e_{i}}^{+}B_{e_{3}}+B_{e_{3}}^{+}B_{e_{i}})]\cdot
\mathbf{E}(t) \label{eq:Heigen}%
\end{align}
Here the interaction with the bath is given by
\begin{equation}
\hbar\left(
\begin{array}
[c]{cc}%
\alpha_{e_{1}} & \alpha_{e_{1}e_{2}}\\
\alpha_{e_{2}e_{1}} & \alpha_{e_{2}}%
\end{array}
\right)  =U^{-1}H_{eb}U=\hbar\left(
\begin{array}
[c]{cc}%
\alpha_{1}\cos^{2}\theta+\alpha_{2}\sin^{2}\theta & \frac{1}{2}(\alpha
_{2}-\alpha_{1})\sin2\theta\\
\frac{1}{2}(\alpha_{2}-\alpha_{1})\sin2\theta & \alpha_{1}\sin^{2}%
\theta+\alpha_{2}\cos^{2}\theta
\end{array}
\right)  \label{eq:1exciton_coordinate}%
\end{equation}
$\allowbreak\allowbreak$and%
\begin{equation}
\alpha_{e_{3}}=\alpha_{1}+\alpha_{2}, \label{eq:2exciton_coordinate}%
\end{equation}
for the single-exciton and two-exciton states, respectively.
Eqs.(\ref{eq:1exciton_coordinate}) and (\ref{eq:2exciton_coordinate}) define
the fluctuating parts of the single-exciton and two-exciton state transition frequencies.

Consider various correlation functions. Assuming that baths acting on
different chromophores are uncorrelated%

\begin{equation}
\langle\alpha_{m}(t)\alpha_{n}(0)\rangle=0\text{ for }m\neq n
\label{eq:bath_uncorrelation}%
\end{equation}
and that the site energy fluctuation correlation functions are identical for
the two monomers \cite{Fle99,Muk04ChemRev}, we get%

\begin{align}
\langle\alpha_{e_{1}}(t)\alpha_{e_{1}}(0)\rangle &  =\langle\alpha_{e_{2}%
}(t)\alpha_{e_{2}}(0)\rangle=\hbar^{-2}K(t)(\cos^{4}\theta+\sin^{4}%
\theta),\nonumber\\
\langle\alpha_{e_{3}}(t)\alpha_{e_{3}}(0)\rangle &  =2\hbar^{-2}K(t)
\label{eq:correlations}%
\end{align}
where $K(t)=\hbar^{-2}\langle\alpha_{1}(t)\alpha_{1}(0)\rangle=\hbar
^{-2}\langle\alpha_{2}(t)\alpha_{2}(0)\rangle\equiv\hbar^{-2}\langle
\bar{\alpha}(t)\bar{\alpha}(0)\rangle$. The further calculations simplify
considerably if the off-diagonal part of the interaction with the bath in the
exciton representation $\alpha_{e_{1}e_{2}}=\alpha_{e_{2}e_{1}}$ in
Eq.(\ref{eq:1exciton_coordinate}) can be neglected. This approximation is
discussed in Refs.\cite{Fle99,Muk97}.

The correlation function $K(t)$ can be represented as the Fourier transform of
the power spectrum $\Phi(\omega)$ of $\hbar\alpha_{1}$($=\hbar\alpha_{2}$)
\cite{Fai03AMPS}%

\[
K(t)=\int_{-\infty}^{\infty}d\omega\Phi(\omega)\exp(i\omega t)
\]
where
\begin{equation}
\Phi(-\omega)=\Phi(\omega)\exp(-\beta\hbar\omega) \label{eq:Fi(-omega)}%
\end{equation}
Using Eq.(\ref{eq:Fi(-omega)}), the real and imaginary parts of
$K(t)=K^{\prime}(t)+iK^{\prime\prime}(t)$ can be written as
\begin{align*}
K^{\prime}(t)  &  =\int_{0}^{\infty}d\omega\Phi(\omega)[1+\exp(-\beta
\hbar\omega)]\cos\omega t\\
K^{\prime\prime}(t)  &  =\int_{0}^{\infty}d\omega\Phi(\omega)[1-\exp
(-\beta\hbar\omega)]\sin\omega t
\end{align*}
In the high temperature limit one get%

\begin{align*}
K^{\prime}(t)  &  =2\int_{0}^{\infty}d\omega\Phi(\omega)\cos\omega t\\
K^{\prime\prime}(t)  &  =\hbar\beta\int_{0}^{\infty}d\omega\Phi(\omega
)\omega\sin\omega t
\end{align*}
where $K(0)=K^{\prime}(0)=2\int_{0}^{\infty}d\omega\Phi(\omega)=\hbar
^{2}\sigma_{2}=\hbar\omega_{St}\beta^{-1}$; $\sigma_{2}$ and $\omega_{St}$ are
a second central moment and the Stokes shift of the equilibrium absorption and
luminescence spectra, respectively, for each monomer.

Similar to Sec.\ref{sec:equations}, we will consider $\bar{\alpha}=-u/\hbar$
as a stochastic Gaussian variable with the correlation function $\langle
\bar{\alpha}(t)\bar{\alpha}(0)\rangle=\sigma_{2}\exp(-|t|/\tau_{s})$
corresponding to the Gaussian-Markovian process. In this case the
Fokker--Planck operators for the excited state of each monomer has the
following form%

\begin{equation}
L_{m}=\tau_{s}^{-1}\left(  \frac{\partial^{2}}{\partial x^{2}}+\left(
x-x_{m}\right)  \frac{\partial}{\partial x}+1\right)  \label{eq:Lmonomer}%
\end{equation}
where $x=q\tilde{\omega}\sqrt{\beta}=\bar{\alpha}/\sqrt{\sigma_{2}}$ is a
dimensionless generalized coordinate. Bearing in mind
Eqs.(\ref{eq:correlations}), the Fokker--Planck operators for the eigenstates
$|j\rangle=$ $|0\rangle$,$|e_{i}\rangle$ of the exciton Hamiltonian can be
written by Eq.(\ref{eq:Ljj}) where $x_{0}=0$, $x_{e_{1}}=x_{e_{2}}=x_{m}%
(\cos^{4}\theta+\sin^{4}\theta)$ and $x_{e_{3}}=2x_{m}$. The corresponding
transition frequencies at the equilibrium nuclear coordinate of the ground
electronic state are defined by Eqs.(\ref{eq:1exciton_eigenvalues}) and
(\ref{eq:2exciton_eigenvalue}).

Consider a homodimer complex consisting of identical molecules with
$\bar{\Omega}_{1}=\bar{\Omega}_{2}\equiv\bar{\Omega}$ and $D_{1}=D_{2}\equiv
D.$ For this case, using Eqs.(\ref{eq:tan2teta}), (\ref{eq:one_exciton}),
(\ref{eq:1exciton_eigenvalues}), (\ref{eq:2exciton_eigenvalue}) and
(\ref{eq:dipole_between1exciton_2exciton}), we obtain $\theta=\pi/4$,%

\begin{align}
\hbar\bar{\Omega}_{e_{1,2}}  &  =\hbar(\bar{\Omega}\pm J),\text{ }\hbar
\bar{\Omega}_{e_{3}}=2\hbar\bar{\Omega}\label{eq:homodimer}\\
D_{e_{1}}  &  =D_{e_{1}e_{3}}=\sqrt{2}D,\text{ }D_{e_{2}}=D_{e_{2}e_{3}%
}=0\nonumber
\end{align}
We thus need to consider only three states: $|0\rangle$, $|e_{1}\rangle$ and
$|e_{3}\rangle$, since state $|e_{2}\rangle$ is not excited with light.
Letting $|1\rangle$, $|2\rangle$ and $|3\rangle$ represent $|0\rangle$,
$|e_{1}\rangle$ and $|e_{3}\rangle$, respectively, we arrive at a three-state
system considered above where $\omega_{21}=\bar{\Omega}+J,$ $\omega_{31}%
=2\bar{\Omega},$ $D_{21}=D_{32}=\sqrt{2}D,$ $x_{1}=0$, $x_{2}=\frac{1}{2}%
x_{m}$, $x_{3}=2x_{m}$.%

\begin{figure}
[ptb]
\begin{center}
\includegraphics[
height=2.0409in,
width=3.4122in
]%
{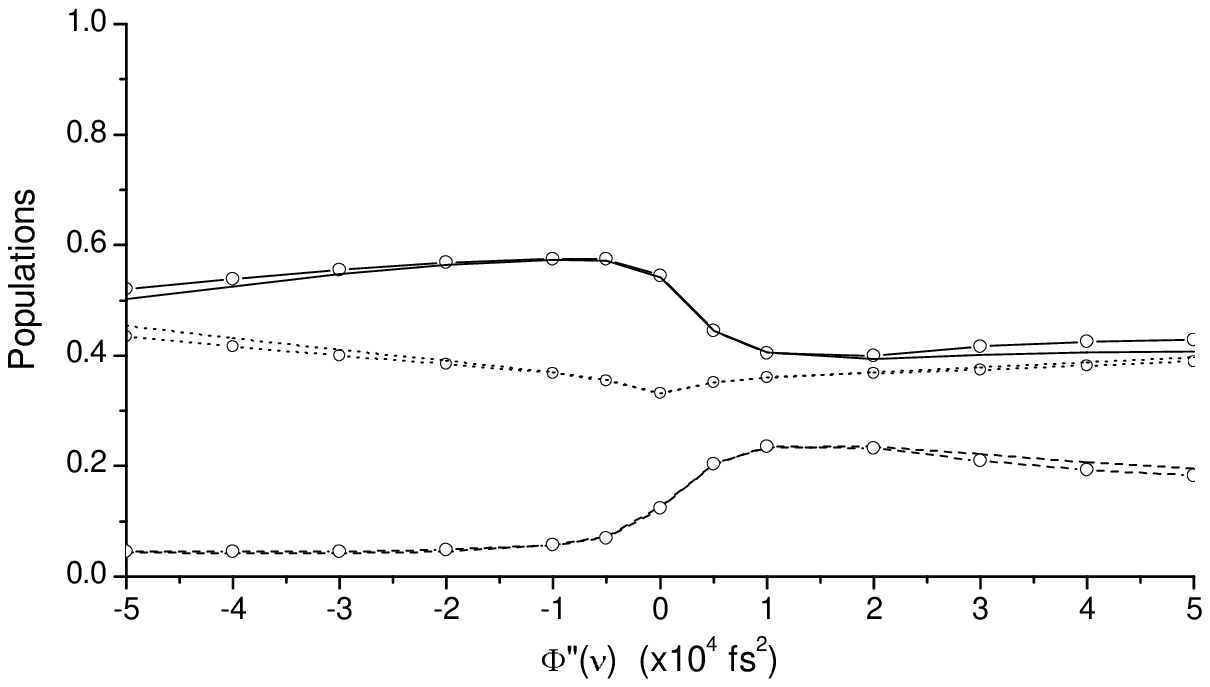}%
\caption{Populations of the ground (dotted line), single- (solid line) and
two-exciton (dashed line) states of a homodimer complex after the completion
of the pulse action as functions of $\Phi"(\nu)$ for $J=-300$ $cm^{-1}$ ($J<0$
- J-aggregate), $Q^{\prime}=2.9$, $t_{p0}=10$ $fs$, $\tau_{s}=100$ $fs$. The
partial\ relaxation and the total \ models - lines with and without
hollow\ circles, respectively. }%
\label{fig:dimer}%
\end{center}
\end{figure}

Fig.\ref{fig:dimer} shows populations of single and two-exciton states after
the excitation with a linear chirped pulse, Eqs.(\ref{eq:gausspulse}) and
(\ref{eq:deltamu}), as functions of $\Phi^{\prime\prime}(\nu)$. Here the
one-photon resonance for Franck-Condon transition $1\rightarrow2$ occurs at
the pulse maximum, i.e. $\omega=\omega_{21}=\bar{\Omega}+J$, and the Stokes
shift of the equilibrium absorption and luminescence spectra for each monomer
is equal to $\omega_{st}^{mon}=400$ $cm^{-1}$. Fig.\ref{fig:dimer} also
contrasts calculations using the total model (lines without hollow\ circles)
with those of the partial relaxation model
when\ only\ diagonal\ matrix\ elements of the density
matrix\ undergo\ diffusion (lines with hollow\ circles). Fig.\ref{fig:dimer}
shows a good agreement between the calculation results for both models.

Furthermore, one can see strong suppressing the population of the two-exciton
state for negatively chirped (NC) pulse excitation. As a matter of fact, one
can suppress or enhance two-exciton processes using positively or NC pulses.
Our calculations (see table below) show twofold benefits of NC pulse
excitation ($\Phi^{\prime\prime}=-10^{4}$ $fs^{2}$) with respect to the
transform limited pulse ($\Phi^{\prime\prime}=0$) of the same duration
($t_{p}=71$ $fs$) and energy tuned to one-exciton transition: the population
transfer to the single exciton state is larger, and that to the two-exciton
state is smaller.

$%
\begin{array}
[c]{ccc}%
\begin{array}
[c]{c}%
\text{Populations after the }\\
\text{completion of pulse action}%
\end{array}
&
\begin{array}
[c]{c}%
\text{Transform limited}\\
\text{pulse (}\Phi^{\prime\prime}=0\text{, }t_{p}=71fs\text{) }%
\end{array}
& \text{NC pulse }(\Phi^{\prime\prime}=-10^{4}fs^{2}\text{, }t_{p}=71fs)\\
n_{2} & 0.317 & 0.573\\
n_{3} & 0.208 & 0.057
\end{array}
$

It is worthy to note good selective properties of chirped pulses, bearing in
mind strong overlapping Franck-Condon transitions $1\rightarrow2$,
$\omega_{21}$, and $2\rightarrow3$, $\omega_{32}$. Really, the corresponding
frequencies differ by $\omega_{32}-\omega_{21}=-2J-\frac{3}{4}\omega
_{st}^{mon}$ for the model under consideration that comes to $\omega
_{32}-\omega_{21}=300$ $cm^{-1}$ for the used values of parameters. On the
other hand, the bandwidth of the absorption spectrum at half maximum for
transition $2\rightarrow3$ comes to $\Delta\omega=2\sqrt{2\ln2\sigma_{2s}%
^{23}}\approx1024$ $cm^{-1}$ that is larger than $\omega_{32}-\omega_{21}$.
Here $\sigma_{2s}^{23}=(\hbar\beta)^{-1}\omega_{st}^{23}$ is the LF vibration
contribution to a second central moment of an absorption spectrum for
transition $2\rightarrow3$ and $\omega_{st}^{23}=(\hbar\beta)^{-1}(x_{3}%
-x_{2})^{2}=\frac{9}{4}\omega_{st}^{mon}$\ is the corresponding Stokes shift.

This issue can be understandable in terms of the competition between
sequential and direct paths in a two-photon transition \cite{Gir03}. Consider
a three-level atomic ladder system in the absence of relaxation with close
transition frequencies $\omega_{21}\approx\omega_{32}$ where $\omega_{21}$ can
be associated with one-exciton excitation and frequency $\omega_{31}$ - with
two-exciton excitation.The system is affected by one phase modulated pulse of
carrier frequency $\omega$, Eqs.(\ref{eq:field}), (\ref{eq:gausspulse}) and
(\ref{eq:deltamu}). In the Appendix we have calculated the excited-state
amplitude $a_{3}$ due to two-photon transition $1\rightarrow3$ involving a
nearly resonant intermediate level $2$ for such system. Amplitude
$a_{3}=a_{TP}+a_{S}$ consists of two contributions. The first one $a_{TP}%
$\ corresponds exactly to that of the nonresonant two-photon transition. This
contribution $a_{TP}\thicksim1/|\Phi^{\prime\prime}(\omega)|$, and it is small
for strongly chirped pulses \cite{Fai00CPL}%

\begin{equation}
2|\Phi^{\prime\prime}(\omega)|\gg\tau_{p0}^{2} \label{eq:strongchirp}%
\end{equation}
This result has a clear physical meaning. The point is that the phase
structure (chirp) of the pulse determines \textit{the temporal ordering} of
its different frequency components. For a strongly chirped pulse when a pulse
duration is much larger than that of the corresponding transform-limited one,
one can ascribe to different instants of time the corresponding frequencies
\cite{Fai00CPL}. As a matter of fact, in the case under consideration
different frequency components of the field are determined via values of the
instantaneous pulse frequency $\omega(t)$ for different instants of time.
Therefore, only a small part of the whole pulse spectrum directly excites the
two-photon resonance.

The second contribution is given by \cite{Gir03}%

\begin{equation}
a_{S}=-\frac{D_{32}D_{21}\pi}{2\hbar^{2}}E(\omega_{21})E(\omega_{32}%
)\{1-sgn[(\omega_{21}-\omega_{32})\Phi^{\prime\prime}\left(  \omega\right)
]\} \label{eq:a_3final}%
\end{equation}
where $E(\tilde{\omega})$ is the Fourier transform of the positive frequency
components of the field amplitude $\mathcal{E}\left(  t\right)  \exp
[i\varphi_{i}\left(  t\right)  ]$. The consideration of the Appendix enables
us to extend the results of Ref.\cite{Gir03} to non-zero two-photon detuning
$\Omega_{2}=\omega_{31}-2\omega\neq0$. Eq.(\ref{eq:a_3final}) describes a
sequential process, the contribution of which is a steplike function. This
process can be suppressed when the pulse frequencies arrive in
counter-intuitive order ($\omega_{32}$ before $\omega_{21}$) that occurs in
our simulations of a J-aggregate for NC excitation. Fig.\ref{fig:dimer} and
the table above show that the selective properties of chirped pulses under
discussion are conserved on strong field excitation and for broad transitions.
The selective excitation of single and two-exciton states can be used for
preparation of initial states for nonlinear spectroscopy based on pulse
shaping \cite{Nel04,Nel05}.

\section{Strong interaction and STIRAP}

\label{sec:STIRAP}

The three-state system under discussion enables us to consider STIRAP as well.
STIRAP in molecules in solution was studied in Refs.\cite{Rice02}, where the
solvent fluctuations were represented as a Gaussian random process, and in
Ref.\cite{Geva03}, where the system-bath coupling was taken to be weak in the
sense that the relaxation times were long in comparison to the bath
correlation time, $\tau_{c}$. Intense fields were shown in Ref.\cite{Geva03}
to effectively slow down the dephasing when the energetic distance between the
dressed (adiabatic) states exceeds $1/\tau_{c}$. The point of the last paper
is that in contrast to usual undressed states, which intersect, the dressed
(adiabatic)\ states do not intersect. Therefore, the spectral density of the
relaxation induced noise, which has a maximum at zero frequency, strongly
diminishes for frequencies corresponding to the light-induced gap between
dressed\ states, resulting in suppressing pure dephasing between the
dressed\ states. In this section we show that this conclusion holds also for
non-Markovian relaxation when the system-bath interaction is not weak and,
therefore, can not be characterized only by $\tau_{c}$.

In the rotating wave approximation the Schr\"{o}dinger equations for STIRAP in
$\Lambda$-configuration can be written as follows%

\begin{equation}
i\hbar\frac{d}{dt}\left(
\begin{array}
[c]{c}%
a_{1}\\
a_{2}\\
a_{3}%
\end{array}
\right)  =\left(
\begin{array}
[c]{ccc}%
U_{1}^{\prime} & -\hbar\Omega_{1}/2 & 0\\
-\hbar\Omega_{1}/2 & U_{2} & -\hbar\Omega_{2}/2\\
0 & -\hbar\Omega_{2}/2 & U_{3}^{\prime}%
\end{array}
\right)  \left(
\begin{array}
[c]{c}%
a_{1}\\
a_{2}\\
a_{3}%
\end{array}
\right)  \label{eq:SchrSTIRAP}%
\end{equation}
where $U_{1}^{\prime}=U_{1}+\hbar\omega_{1}$ and $U_{3}^{\prime}=U_{3}%
+\hbar\omega_{2}$ are \textquotedblright photonic
replications\textquotedblright\ of effective parabolic potentials $U_{1}(x)$
and $U_{3}(x)$ (Eq.(\ref{eq:Uj})), respectively. We consider the two-photon
resonance condition when $\omega_{1}-\omega_{2}=(E_{3}-E_{1})/\hbar$ and
$x_{1}=x_{3}=0$ that would appear reasonable when $|1\rangle$ and $|3\rangle$
are different vibrational levels of the same electronic state. Then
$U_{1}^{\prime}=U_{3}^{\prime}$.

Adiabatic states $U^{ad}$ corresponding to Eq.(\ref{eq:SchrSTIRAP}) can be
found by equation%

\[
\det\left(
\begin{array}
[c]{ccc}%
U_{1}^{\prime}-U^{ad} & -\hbar\Omega_{1}/2 & 0\\
-\hbar\Omega_{1}/2 & U_{2}-U^{ad} & -\hbar\Omega_{2}/2\\
0 & -\hbar\Omega_{2}/2 & U_{3}^{\prime}-U^{ad}%
\end{array}
\right)  =0
\]
This gives the following adiabatic states
\begin{align}
U_{0}^{ad}  &  =U_{1}^{\prime}=U_{3}^{\prime}\nonumber\\
U_{\pm}^{ad}  &  =\frac{1}{2}(U_{2}+U_{1}^{\prime})\pm\frac{1}{2}\sqrt
{(U_{2}-U_{1}^{\prime})^{2}+\hbar^{2}(\Omega_{1}^{2}+\Omega_{2}^{2})}
\label{eq:ad}%
\end{align}
One can see that initial $U_{1}^{\prime}$ and final $U_{3}^{\prime}$ diabatic
states coincide with one of adiabatis states $U_{0}^{ad}$. For strong
interaction the last will be well separated from other adiabatic states
$U_{\pm}^{ad}$ due to avoided crossing. Therefore, during STIRAP the system
will remain in the same adiabatic state $U_{0}^{ad},$ which is $U_{1}^{\prime
}$ for $t=-\infty$ and $U_{3}^{\prime}$ for $t=+\infty$. Its evolution due to
relaxation stimulated by LF vibrations can be described by the corresponding
Fokker-Planck operator $L_{0}^{ad}=L_{1,3}=\tau_{s}^{-1}\left(  \frac
{\partial^{2}}{\partial x^{2}}+x\frac{\partial}{\partial x}+1\right)  $
describing diffusion in adiabatic potential $U_{0}^{ad}=U_{1}^{\prime}%
=U_{3}^{\prime}$. This means that during transition $1\rightarrow3$ the system
motion along a generalized coordinate $x$ does not change. In other words,
such a transition will not be accompanied by pure dephasing. This conclusion
is a generalization of the previous result \cite{Geva03} relative to slowing
down the dephasing in strong fields, which was obtained for weak system-bath
interaction, to non-Markovian relaxation.

\section{Conclusion}

\label{sec:conclusion}

In this work we have studied the influence of ESA and two-exciton processes on
a coherent population transfer with intense ultrashort chirped pulses in
molecular systems in solution. An unified treatment of ARP in such systems has
been developed using{ a three-state electronic system with relaxation treated
as a diffusion on electronic potential energy surfaces. We believe that such a
simple model properly describes the main relaxation processes related to
overdamped motions occurring in large molecules in solutions. }

Our calculations show that even with fast relaxation of a higher singlet state
$S_{n}$ ($n>1$) back to $S_{1}$, ESA has a profound effect on coherent
population transfer in complex molecules that necessitates a more accurate
interpretation of the corresponding experimental data. In the absence of
$S_{n}\rightarrow S_{1}$ relaxation, the population of state $|3\rangle$,
$n_{3}$, strongly decreases when the chirp rate in the frequency domain
$|\Phi^{\prime\prime}(\nu)|$ increases. In order to appreciate the physical
mechanism for such behavior, an approach to the total model - {the
relaxation-free model -} {was invoked. A comparison between }the total model
behavior and that of the {relaxation-free model\ has shown that }relaxation is
responsible for strong decreasing $n_{3}$ as a function of $\Phi^{\prime
\prime}(\nu)$ in spite of meeting adiabatic criteria for both transitions
$1\rightarrow2$ and $2\rightarrow3$ separately. {By this means }usual criteria
for ARP in a two-state system must be revised for a three-state system.

To clarify this issue, we have developed a simple and physically clear model
for ARP with a linear chirped pulse in molecules with three electronic states
in solution. The relaxation effects were considered in the framework of the LZ
calculations putting in a third level generalized for random crossing of
levels. The model has enabled us to obtain a simple formula for $n_{3}$,
Eq.(\ref{eq:n3_mu+_3}), which is in excellent agreement with numerical
calculations. In addition, the model gives us an extra criterion for coherent
population transfer to those we have obtained before for a two-state system
\cite{Fai04JCP}. New criterion, Eq.(\ref{eq:counter-movement}), implies
conservation of the \textquotedblleft counter-movement\textquotedblright\ of
the \textquotedblleft photonic repetitions\textquotedblright\ of states $1$
and $3$, in spite of random crossing of levels.

Furthermore, we also applied our model to a molecular dimer consisting from
two-level chromophores. A strong suppressing of two-exciton state population
for NC pulse excitation of a J-aggregate has been demonstrated. We have shown
that one can suppress or enhance two-exciton processes using positively or NC
pulses. As a matter of fact, a method for quantum control of two-exciton
states has been proposed. Our calculations show good selective properties of
chirped pulses in spite of strong overlapping transitions related to the
excitation of single- and two-exciton states.

In the light of the limits \cite{Fra99,Goy01} imposed on
Eqs.(\ref{eq:rho12tilda}) and (\ref{eq:rho23tilda}) for nondiagonal elements
of the density matrix for the total model, we used a semiclassical (Lax)
approximation (Eq.(\ref{eq:rho12tilda_without})) (the partial relaxation
model). The latter offers a particular advantage over the total model. The
point is that the partial relaxation model can be derived not assuming the
standard adiabatic elimination of the momentum for the non-diagonal density
matrix, which is incorrect in the \textquotedblright slow
modulation\textquotedblright\ limit \cite{Pol03}. A good agreement between
calculation results for the partial relaxation and the total models in the
slow modulation limit (see Figs.\ref{fig:ESA_models} and \ref{fig:dimer})
shows that a specific form of the relaxation term in the equations for
nondiagonal elements of the density matrix $\tilde{\rho}_{12}(x,t)$ and
$\tilde{\rho}_{23}(x,t)$ is unimportant. By this means the limits imposed on
the last equation \cite{Fra99,Goy01} are of no practical importance for the
problem under consideration in the slow modulation limit. This issue can be
explained as follows. Our previous simulations \cite{Fai02JCP_2} show that in
spite of a quite different behavior of the coherences (nondiagonal density
matrix elements) for the partial relaxation and the total models, their
population wave packets $\rho_{jj}\left(  x,t\right)  $ behave much like.
Since we are interested in the populations of the electronic states
$n_{j}=\int\rho_{jj}\left(  x,t\right)  dx$ only, which are integrals of
$\rho_{jj}\left(  x,t\right)  $ over $x$, the distinctions between the two
models under discussion become minimal.

In conclusion, we have also demonstrated slowing down the pure dephasing on
STIRAP in strong fields when the system-bath interaction is not weak
(non-Markovian relaxation).

\textbf{Acknowledgement}

This work was supported by the Ministry of absorption of Israel.

{\Huge Appendix}

Consider a three-level system $E_{1}<E_{2}<E_{3}$ with close transition
frequencies $\omega_{21}\approx\omega_{32}$ where $\omega_{21}$ can be
associated with a single-exciton excitation and frequency $\omega_{31}$ - with
two-exciton excitation. The system is affected by one phase modulated pulse of
carrier frequency $\omega$, Eq.(\ref{eq:field}). The excited-state amplitude
for a two-photon transition involving a nearly resonant intermediate level,
can be written as \cite{Sil01,Gir03}%

\begin{equation}
a_{3}=-\frac{D_{32}D_{21}}{2\hbar^{2}}\pi\left[  E(\omega_{21})E(\omega
_{32})+\frac{i}{\pi}P\int_{-\infty}^{\infty}d\Omega\frac{E(\Omega
+\omega)E(\Omega_{2}-\Omega+\omega)}{\Omega-(\omega_{21}-\omega)}\right]
\label{eq:a_3}%
\end{equation}
where $E(\tilde{\omega})$ is the Fourier transform of the positive frequency
components of the field amplitude $\mathcal{E}\left(  t\right)  \exp
[i\varphi_{i}\left(  t\right)  ]$, $\Omega=\tilde{\omega}-\omega$, $P$ is the
principal Cauchy value, $\Omega_{2}=\omega_{31}-2\omega$ is the two-photon
detuning. For linear chirped excitation, Eqs.(\ref{eq:gausspulse}) and
(\ref{eq:deltamu}), $E(\tilde{\omega})$ is given by%

\begin{equation}
E(\tilde{\omega})=\sqrt{\pi}\mathcal{E}_{0}\tau_{p0}\exp\{-\frac{1}{2}%
\Omega^{2}[\tau_{p0}^{2}/2-i\Phi^{\prime\prime}(\omega)]\} \label{eq:LinChirp}%
\end{equation}
Using Eq.(\ref{eq:LinChirp}) and introducing a new variable $z=\Omega
-\Omega_{2}/2$, Eq.(\ref{eq:a_3}) can be written as%

\begin{align}
a_{3}  &  =-\frac{D_{32}D_{21}\pi^{2}(\mathcal{E}_{0}\tau_{p0})^{2}}%
{2\hbar^{2}}\{\exp[-\frac{1}{2}(\delta^{2}+(\Omega_{2}-\delta)^{2})(\tau
_{p0}^{2}/2-i\Phi^{\prime\prime}(\omega))]\nonumber\\
&  +\exp[-\frac{1}{4}\Omega_{2}^{2}(\tau_{p0}^{2}/2-i\Phi^{\prime\prime
}(\omega))]\frac{i}{\pi}P\int_{-\infty}^{\infty}dz\frac{\exp[-z^{2}(\tau
_{p0}^{2}/2-i\Phi^{\prime\prime}(\omega))]}{z-(\delta-\Omega_{2}/2)}\}
\label{eq:a_3a}%
\end{align}
where $\delta=\omega_{21}-\omega$ is one-photon detuning. The integral on the
right-hand side of Eq.(\ref{eq:a_3a}) can be evaluated for strongly chirped
pulses \cite{Fai00CPL}, Eq.(\ref{eq:strongchirp}), when a pulse duration is
much larger than that of the corresponding transform-limited one. In this case
two frequency ranges give main contributions to the integral. The first one
results from the method of stationary phase \cite{Fed87}, and it is localized
near the two-photon resonance $z=\omega-\omega_{31}/2=0$ in the small range
$\Delta\omega\sim1/\sqrt{|\Phi^{\prime\prime}(\omega)|}$. In this case only a
small part $\Delta\omega\sim1/\sqrt{|\Phi^{\prime\prime}\left(  \omega\right)
|}$ of the whole pulse spectrum $\Delta\omega_{pulse}=4/\tau_{p0}$ directly
excites the two-photon resonance, and the corresponding contribution
$\sim1/\sqrt{|\Phi^{\prime\prime}(\omega)|}$ is small due to
Eq.(\ref{eq:strongchirp}).

The second contribution to the integral is located near $z=\delta-\Omega
_{2}/2$ and it is due to the pole at the real axes. This contribution is given
by Eq.(\ref{eq:a_3final}) of Sec.\ref{sec:excitons}.


\end{document}